\documentclass[sigconf, nonacm]{acmart}

\usepackage[most]{tcolorbox}
\usepackage{enumitem}

\begin{document}

\title{Revisiting Feedback Models for HyDE}

\author{Nour Jedidi}
\affiliation{
    \institution{University of Waterloo}
    \city{Waterloo}
    \state{ON}
    \country{Canada}}

\email{njedidi@uwaterloo.ca}

\author{Jimmy Lin}
\affiliation{
    \institution{University of Waterloo}
    \city{Waterloo}
    \state{ON}
    \country{Canada}
}
\email{jimmylin@uwaterloo.ca}

\begin{abstract}
Recent approaches that leverage large language models (LLMs) for pseudo-relevance feedback (PRF) have generally not utilized well-established feedback models like Rocchio and RM3 when expanding queries for sparse retrievers like BM25. Instead, they often opt for a simple string concatenation of the query and LLM-generated expansion content. \emph{But is this optimal}? To answer this question, we revisit and systematically evaluate traditional feedback models in the context of HyDE, a popular method that enriches query representations with LLM-generated hypothetical answer documents. Our experiments show that HyDE's effectiveness can be substantially improved when leveraging feedback algorithms such as Rocchio to extract and weight expansion terms, providing a simple way to further enhance the accuracy of LLM-based PRF methods.
\end{abstract}

\maketitle
\pagestyle{empty}
\section{Introduction}

HyDE~\cite{gao2023precise} represents a simple, yet effective method to improve the accuracy of traditional ``bag of words'' retrieval methods such as BM25 by
utilizing large language models (LLMs) to expand queries with hypothetical answer documents. The core concept underlying HyDE is that the knowledge contained within LLMs can be leveraged as a form of \emph{relevance feedback}, helping bridge the vocabulary gap between query representations and those of relevant documents.  This builds upon decades of research in information retrieval (IR) on \emph{pseudo}-relevance feedback (PRF), which traditionally assumed top-ranked documents from a first-stage retriever (like BM25) contain useful expansion terms that can be used for enriching the query representation. Other than obvious distinctions that are simply due to advancements in modeling, the core difference between traditional PRF techniques and LLM-based relevance feedback techniques is solely the \emph{source} of the feedback terms: traditional PRF methods extract feedback terms from top-ranked retrieved documents, whereas HyDE's feedback terms are generated directly by the LLM. 

Despite this seemingly small distinction, the exact mechanism for updating the query representation for BM25 surprisingly differs between the two approaches.  Traditional PRF techniques typically run two phases after an initial set of candidate feedback documents are returned~\cite{carpineto2002improving}. The first phase consists of selecting the terms to be included in the expanded query, $q_{new}$. The second phase consists of properly weighting these selected terms,  along with those in
the original query,  for $q_{new}$.  On the other hand,  LLM relevance feedback methods like HyDE and its variants have generally opted for simple concatenation strategies that directly feed the LLMs generated text and the original query into BM25 with minimal processing. 

Motivated by this gap, we ask: \emph{is it justified to bypass the traditional two-phase approach or are current methods for updating the query representation with HyDE suboptimal?} To answer this question, we revisit feedback models for scoring and weighting expansion terms generated by HyDE, an area that has been largely ignored by recent LLM-based relevance feedback research. In this paper, we focus primarily on feedback models that remain widely used today, such as Rocchio~\cite{rocchio1971relevance} and RM3~\cite{abdul2004umass}. 

To evaluate whether traditional feedback models can improve upon current approaches, we compare them against string concatenation methods across 14 retrieval datasets. Our results demonstrate that current methods used to improve query representations for BM25 with HyDE are suboptimal, especially when applied to low-resource retrieval tasks. For example, by simply utilizing the Rocchio feedback algorithm, HyDE’s effectiveness can improve by as much as 1.4 points (4.2\%) on average versus a top-performing string concatenation approach.

Our contributions are as follows. First, we run a comprehensive evaluation of HyDE with different feedback mechanisms. While recent work has primarily focused on how to improve the expansions that the LLM produces, relatively little effort has been focused on evaluating different ways an LLMs generated feedback can be leveraged to update the query representation for BM25. To the best of our knowledge, our study represents the first to properly compare different feedback mechanisms, spanning new and traditional techniques, in the context of LLM generated feedback.  Second, our results demonstrate that the vast literature on feedback models, which has been mostly overlooked by recent methods, remains highly relevant and can provide substantial improvements in the effectiveness of HyDE with BM25. Third, to facilitate future research on leveraging feedback models with LLM generated expansions, we open-source our code at \url{https://github.com/nourj98/hyde-feedback}.
\section{Feedback Models}
\label{sec: feedback_methods}
Given a query, $q$, HyDE first prompts an LLM to generate a hypothetical answer document and samples $N$ variations, $e=\{\hat{d}_1, \dots, \hat{d}_N\}$. The terms present in these hypothetical documents form the basis of what is subsequently fed into the feedback models that (1) select the terms from $e$ to include in the updated query, $q_{new}$, and (2) generate the query term-weights for $q_{new}$. In this section, we overview each of these steps in more detail.  

\subsection{Selecting Feedback Terms}
\label{sec: term_selection}
Once HyDE has generated  $e$, we generate a term-frequency vector, $f(\hat{d}_i)$, for each of the feedback documents in $e$. We then directly follow the process  implemented in Anserini~\cite{yang2017anserini}, which, at a high-level, first filters out common corpus terms from each feedback vector, keeping only the terms which occur in less than 10\% of the corpus documents. From these filtered vectors, $\tilde f(\hat d_i)$, candidate expansion terms are then ranked by the sum of their normalized term frequencies across each of $\tilde f(\hat d_i)$ and, from this, $\tilde f(\hat d_i)$ is further pruned to only include the top-$k$ ranked feedback terms~\cite{carpineto2002improving}.\footnote{The specifics of this selection procedure can be different based on the feedback model.} 

Now that we have a set of feedback terms for each of the HyDE generated documents, we next discuss the different feedback models considered in our evaluation for generating $q_{new}$.

\subsection{HyDE Query Update: Average Vector}

The first feedback model we consider is a reformulation of HyDE's original query update, which was originally designed for dense vectors, to the bag-of-words vector space. The HyDE query update takes a simple average of the query representation and the representations of the hypothetical answer documents. In particular, it considers the query as an additional feedback document in $e$, i.e., $e_{\text{HyDE}}=\{q, \hat{d}_1, \dots, \hat{d}_N\}$. Adapted to the bag-of-words space, the weight of a term $w_{t, q_{new}}$ in  $q_{new}$ is computed as follows:

\begin{equation}
    w_{t, q_{\text{new}}}
    = \frac{1}{N + 1}
    \sum_{d_{i} \in e_{\text{HyDE}}} 
        \tilde f(d_i)[t]
    \label{eq:HyDE_with_query}
\end{equation}

\noindent
Here, $\tilde f(d_i)[t]$ is the normalized frequency of $t$ in $\tilde f(d_i)$. Note that $\tilde f$ is computed as described in Section~\ref{sec: term_selection}, however, for $q$, we do not filter out common corpus terms. 

At a high-level, this represents a term-weight as its average across the query and HyDE feedback documents. We note that this, by itself, would return similar rankings by BM25 to the following string concatenation,  $q_{new} = \text{Concat}\left(q , e\right)$ if no selection of feedback terms is performed.\footnote{This assumes that BM25 creates the query vector using a term-frequency vector, as described in~\cite{ge2025lighting}, which would only cause BM25 similarity scores to differ by a multiplicative constant due to leveraging a sum rather than a mean term-weight.} As such, this feedback method allows us to evaluate how useful the \emph{term selection} phase is.  

\subsection{Traditional Models: Rocchio and RM3}

We next consider Rocchio's algorithm, which is a more generalized version of the  HyDE vector update that instead controls the weight of the query vector and feedback vectors using parameters, $\alpha$ and $\beta$, rather than an equal weighting: 

\begin{equation}
    w_{t, q_{\text{new}}}
    = \alpha f(q)[t] + \frac{\beta}{N} \sum_{\hat{d}_i \in e}  \tilde f(\hat d_i)[t]
    \label{eq:HyDE}
\end{equation}

\noindent
Here, $f(q)$ is the normalized term-frequency vector of $q$, and $f(q)[t]$ is the normalized frequency of $t$ in $f(q)$.

Lastly, we consider the RM3 feedback approach~\cite{abdul2004umass}. Given feedback documents, RM3 estimates a relevance model that computes the probability that a word is observed in a document relevant to the query~\cite{lavrenko2017relevance}. In our case, this probability is estimated based on the hypothetical documents in $e$. This relevance model is then linearly interpolated with the probability of the term given the original query. Using our above framework, the term weight would be computed as follows:

\begin{equation}
    w_{t, q_{\text{new}}}
    = \lambda P(t|q) + (1 - \lambda) \sum_{\hat{d}_i \in e} P(t|\hat{d}_i)
    \label{eq:HyDE}
\end{equation}

\noindent
The exact mechanism for how to compute $P(t|\hat{d}_i)$ is beyond the scope of this paper. We refer readers to Abdul-Jaleel et al.~\cite{abdul2004umass} and to the RM3 implementation in Anserini~\cite{yang2017anserini}, which we adopted in our experiments.\footnote{\href{https://github.com/castorini/anserini/blob/master/src/main/java/io/anserini/rerank/lib/Rm3Reranker.java}{See here for Anserini implementation of RM3 with BM25. } }

Once the weights for each feedback term in $q_{\text{new}}$ have been computed, we construct custom Lucene queries in which each term's boost is set to $w_{t, q_{\text{new}}}$.

\begin{table*}[t!]
\centering
\caption{Main results (Recall@20) across MS MARCO and BEIR.}
\resizebox{\textwidth}{!}{%
\begin{tabular}{l|ccccc|ccccccccc|c|c}
\toprule
\multicolumn{1}{c}{}  & \multicolumn{5}{c}{\textbf{MS MARCO}} & \multicolumn{10}{c}{\textbf{BEIR}} \\
\cmidrule(lr){2-6} \cmidrule(lr){7-16}
& DL19 & DL20 & DL21 & DL22 & DL23 & Covid 	& News &	SciFact &	FiQA &	DBPedia &	NFCorpus &	Robust04 &	SCIDOCS & ArguAna & BEIR (Avg.) & All (Avg.) \\
\midrule
BM25  & 27.0  &	32.6  &	14.7  &	4.4  &	12.5  &	3.0  &	20.8  &	85.2	 & 37.1	  & 28.3  &	18.0	 & 20.0  &	20.6  &	79.1  &	34.7 & 28.8 \\
\quad w/ Avg Vector & \bf 30.1	 & 35.2	 & 17.4	 & 5.0 &	12.5 &	3.0 &	24.3 &	86.5 &	31.5 &	28.3 &	21.7 &	19.3 &	21.6 &	79.7 & 35.1 & 29.7\\
\quad w/ RM3  & 28.9	 & \bf 36.3  &	\bf 17.5  &	4.9	  & 12.6  &	\bf 3.3	 & 24.9  &	86.9  &	\bf 35.4  &	28.7  &	22.0  &	\bf 21.0  &	22.0  &	\bf 85.5  &	\bf 36.6  & \bf 30.7 \\
\quad w/ Rocchio & 28.7 &	35.3 &	17.3 &	\bf 5.1 &	\bf 12.9 &	3.1 &	\bf 25.2 &	\bf 87.4 &	33.7	 & \bf 29.1	& \bf 22.3	& 20.8	& \bf 22.0 &	81.9 & 36.2 & 30.4 \\
\midrule
\multicolumn{7}{l}{\textbf{Qwen2.5-7B}} \\
\midrule
Query2Doc & 34.1 &	44.5 &	18.2 &	10.6 &	\bf 17.1 &	3.6 &	24.2 &	\bf 90.5 &	37.7	 & 31.3 &	21.3 &	21.9 &	21.3 &	80.8 &	37.0 & 32.6 \\
HyDE & & & & & & & & & & & & & & \\
\quad  w/ Naive Concat & 29.5 & 36.9 & 17.6 & 11.1 & 14.6 & 3.6 & 22.4 & 88.9 & 29.2 & 27.4 & 20.2 & 17.6 & 21.0 & 69.9 & 33.3 & 29.3\\
\quad  w/ MuGI Concat & \bf 37.3 & \bf 46.4 & \bf 20.3 & 12.2 & 16.5 & \bf 4.0 & 26.4 & 89.4 & 36.6 & 33.6 & 21.3 & 21.0 & 22.0 & 74.0 & 36.5 & 32.9\\
\quad  w/ Avg Vector & 33.3 & 41.9 & 18.8 & \bf 12.3 & 15.1 & 3.4 & 26.8 & 89.3 & 35.8 & 29.2 & 21.7 & 23.0 & \bf 22.6 & 78.1 & 36.6 & 32.2 \\
\quad  w/ RM3 & 32.8 & 40.1 & 19.5 & 9.1 & 15.1 & 3.7 & 26.4 & 89.3 & \bf 40.6 & 32.3 & 21.2 & 23.3 & 22.2 & 82.4 & 37.9 & 32.7 \\
\quad  w/ Rocchio & 36.9 & 45.5 & 20.3 & 11.7 & 16.5 & 3.9 & \bf 27.0 & 90.5 & 39.9 & \bf 33.9 & \bf 21.8 & \bf 23.7 & 22.4 & \bf 82.9 & \bf 38.4 & \bf 34.0 \\
\midrule
\multicolumn{7}{l}{\textbf{Qwen3-14B}} \\
\midrule
Query2Doc & 32.5 &	44.2 &	20.9 &	10.1 &	15.5 &	3.7 &	26.5 &	90.9 &	38.3 &	32.5 &	21.5 &	22.1 &	22.2 &	81.1 &	37.7 & 33.0 \\
HyDE & & & & & & & & & & & & & & \\
\quad  w/ Naive Concat & 29.2 & 38.7 & 16.9 & 11.4 & 14.8 & 3.4 & 24.8 & 90.1 & 31.5 & 27.3 & 21.2 & 18.0 & 21.1 & 74.8 & 34.7 & 30.2 \\
\quad  w/ MuGI Concat & 37.3 & 46.0 & 22.5 & \bf 12.4 & \bf 16.4 & \bf 3.9 & 25.9 & 90.9 & 36.9 & \bf 34.0 & 21.6 & 21.2 & 22.8 & 76.4 & 37.1 & 33.4 \\
\quad  w/ Avg Vector & 30.7 & 42.5 & 18.1 & 11.8 & 15.3 & 3.3 & 27.8 & \bf 91.1 & 36.3 & 29.2 & \bf 22.3 & 23.1 & \bf 23.7 & 79.8 & 37.4 & 32.5 \\
\quad  w/ RM3 & 33.0 & 40.5 & 20.3 & 8.8 & 14.8 & 3.8 & 27.4 & 90.8 & 41.1 & 32.8 & 21.4 & 24.0 & 22.4 & 83.3 & 38.5 & 33.2\\
\quad  w/ Rocchio & \bf 37.4 & \bf 47.3 & \bf 22.9 & 11.8 & 16.2 & 3.8 & \bf 27.8 & 90.9 & \bf 41.4 & 33.7 & 22.1 & \bf 24.0 & 23.4 & \bf 83.8 & \bf 39.0 & \bf 34.7 \\
\midrule
\multicolumn{7}{l}{\textbf{gpt-oss-20b}} \\
\midrule
Query2Doc & 37.5	& 44.0 &	20.6 &	9.9 &	\bf 15.5 &	3.8 &	24.8 &	87.7 &	39.0 &	31.2 &	20.6 &	21.4 &	21.1 &	80.9 &	36.7 & 32.7\\
HyDE & & & & & & & & & & & & & & \\
\quad  w/ Naive Concat & 28.9 & 38.8 & 18.9 & 8.8 & 14.0 & 3.1 & 24.5 & 91.8 & 31.3 & 26.5 & 20.6 & 18.2 & 14.7 & 72.4 & 33.7 & 29.5 \\
\quad  w/ MuGI Concat & \bf 38.4 & 45.3 & 23.0 & 12.1 & 14.9 & 3.8 & 26.0 & 90.6 & 39.4 & 33.6 & 21.1 & 21.9 & 20.9 & 74.8 & 36.9 & 33.3 \\
\quad  w/ Avg Vector & 32.9 & 42.0 & 19.5 & 10.3 & 13.9 & 3.2 & 28.7 & 91.5 & 37.7 & 28.9 & \bf 22.5 & 24.1 & 21.8 & 78.6 & 37.4 & 32.5\\
\quad  w/ RM3 & 31.8 & 41.4 & 20.4 & 9.2 & 14.9 & \bf 3.9 & 27.0 & 91.0 & \bf 42.1 & 32.4 & 21.6 & 24.5 & 22.2 & 82.5 & 38.6 & 33.2 \\
\quad  w/ Rocchio & 36.9 & \bf 45.8 & \bf 23.5 & \bf 12.3 & 15.4 & 3.8 & \bf 28.1 & \bf 91.9 & 42.0 & \bf 33.8 & 22.1 & \bf 24.6 & \bf 22.7 & \bf 82.5 & \bf 39.1 & \bf 34.7\\

\bottomrule
\end{tabular}
}
\label{tab:msmarco}
\end{table*}

\section{Experimental Setup}
The aim of our experiments is to understand how leveraging the feedback models described in Section ~\ref{sec: feedback_methods} compares to the string-concatenation operation used in recent LLM-based query expansion methods. 

Our HyDE implementation directly follows the approach described in~\cite{gao2023precise}. We sample eight hypothetical documents of up to 512 tokens. When implementing HyDE, we consider three LLMs: Qwen2.5-7B-Instruct~\cite{yang2024qwen2}, Qwen3-14B~\cite{yang2025qwen3}, and gpt-oss-20b~\cite{agarwal2025gpt}.

For the average vector, Rocchio, and RM3 feedback models, we stick to the default hyperparameters used in the literature~\cite{lin2021pyserini}. Namely, for Rocchio, we set $\alpha = 1$ and $\beta = 0.75$. For RM3, we set the query weight, $\lambda$, to 0.5. All feedback models select up to $k=128$ feedback terms to match the number of feedback terms used by Query2Doc~\cite{wang2023query2doc}, one of our string-concatenation baselines.

For the string-concatenation baselines, we compare with a naive concatenation baseline, Query2Doc, and MuGI~\cite{zhang-etal-2024-exploring-best}. Our naive concatenation baseline simply concatenates the query to the HyDE generated documents, $q_{new} = \text{Concat}\left(q , e\right)$. As mentioned, this shares similarity to the average vector feedback approach, but does not perform any selection of the feedback terms, instead using all available terms.  Query2Doc is similar to HyDE in that it expands queries using hypothetical answer documents, but only samples a single hypothetical document of 128 tokens, with a constant query repeat of five, i.e., $q_{new} = \text{Concat}(q, q, q, q, q, \hat{d}_1)$. This comparison allows us to directly study how implementing HyDE with feedback models compares to a more optimal string-concatenation baseline under an approximately equal budget of feedback terms (128). MuGI serves as our \emph{upper bound} on string-concatenation effectiveness, as it leverages all of HyDE's generated documents in its updated query with an adaptive query repeat, rather than a single or constant repeat. MuGI adaptively repeats the query using the following equation:

\begin{equation}
\gamma = \frac{\mathrm{len}(\hat{d}_1) + \mathrm{len}(\hat{d}_2) + \cdots + \mathrm{len}(\hat{d}_n)}{\mathrm{len}(q) \cdot \phi}
\end{equation}

\noindent
As such, $q_{new} = \text{Concat}(q \times \gamma,  e)$. We set $\phi = 5$, following~\cite{zhang-etal-2024-exploring-best}. 

Lastly, we compare against the average vector, Rocchio and RM3 feedback models when using top-ranked BM25 documents rather than HyDE-generated hypothetical documents. This allows us to evaluate how LLM-generated feedback compares to traditional PRF under matched conditions. We directly leverage the same hyperparameters described above for Rocchio and RM3, and use eight feedback documents to match HyDE. We additionally consider BM25 without any query expansion. 

We evaluate all approaches on five web-search datasets from MS MARCO v1 (TREC DL19 and TREC DL20)~\citep{Craswell2020OverviewOT, Craswell2021OverviewOT}, and from MS MARCO v2 (TREC DL21, TREC DL22, and TREC DL23)~\citep{Craswell2021OverviewOT_21, Craswell2022OverviewOT, Craswell2023OverviewOT}. We also evaluate on nine low-resource retrieval datasets from BEIR~\cite{thakur2021beir}. The tasks include news retrieval (TREC-News, Robust04), financial question answering (FiQA), entity retrieval (DBpedia), biomedical IR (TREC-Covid, NFCorpus), fact checking (SciFact), citation prediction (SciDocs) and argument retrieval (ArguAna). For metrics, we report Recall@20.

All retrieval experiments were conducted using the BM25 implementation from Pyserini~\cite{lin2021pyserini} with default parameters. Pyserini additionally provides useful tools to fetch index statistics for generating the feedback document vectors, $\tilde f(\hat d_i)$,  via its IndexReader API and to create custom Lucene queries using its Query Builder API. LLM inference was performed using vLLM~\cite{woosuk2023vllm}. 

\section{Results}

Experimental results on the MS MARCO and BEIR datasets are shown in Table \ref{tab:msmarco}. Below we highlight the key findings:

\begin{itemize}[leftmargin=15pt]

\item \textbf{Applying feedback models to LLM-generated documents is more effective than applying them to top-ranked documents from BM25}. When controlling for the feedback model (Avg Vector, RM3, Rocchio), we find that across LLM families, applying relevance feedback over HyDE documents consistently improves Recall@20. For example, feedback with Avg Vector, RM3, and Rocchio over HyDE (gpt-oss-20b) documents yields a 2.8, 2.5, and 4.3 point improvement, respectively,  compared to feedback with BM25 retrieved documents across all datasets (MS MARCO and BEIR). This trend is stable regardless of the LLM, with similar improvements (2.5, 2, and 3.7 points on average, respectively) when running HyDE using Qwen2.5-7B. Interestingly, we note that while RM3 is less effective than Rocchio when applied to HyDE documents, this is not the case when applied to top-ranked BM25 documents. 

\item \textbf{Simply incorporating a term selection phase as described in Section \ref{sec: term_selection} substantially improves HyDE's effectiveness}. Comparing HyDE (w/ Avg Vector) to HyDE (w/ Naive Concat), we find that the average vector approach is more effective with a 3, 2.3, and 3.1 point improvement across Qwen2.5-7B, Qwen3-14B, and gpt-oss-20b, respectively, on average. As the main difference between HyDE (w/ Avg Vector) and HyDE (w/ Naive Concat)  lies in the term selection step, this demonstrates that it can be helpful to filter out noisy terms from HyDE's feedback documents. 

\item \textbf{Of all the feedback methods, Rocchio demonstrates the strongest effectiveness with HyDE}. Across all LLMs, HyDE (w/ Rocchio) consistently outperforms the other feedback models and string concatenation methods. Compared to HyDE (w/ Avg Vector), which operates in a similar vector space, the crucial difference is that the Rocchio algorithm gives more weight to the query terms. This suggests that the HyDE feedback \emph{under emphasizes} query terms. Additionally, one advantage of the Rocchio algorithm compared to the adaptive query reweighting of MuGI Concat is that the query weight can be adapted linearly with a single parameter rather than being dependent on other factors such as feedback document length. An interesting area for future work would be to investigate how to adaptively reweight the $\alpha$ and $\beta$ parameter for Rocchio using a scheme like that of MuGI.

\item  \textbf{String concatenation methods are competitive on MS MARCO, but are less effective on low-resource BEIR tasks}. Compared to feedback models, Query2Doc and MuGI Concat are generally competitive on the MS MARCO datasets, but are consistently less effective  on BEIR. Taking gpt-oss-20b with RM3 as a representative case, Query2Doc and MuGI Concat beat RM3 on all MS MARCO datasets. However, on BEIR, RM3 beats Query2Doc on all 10 datasets, and MuGI Concat on 8 of the 10. This suggests that updating the query representation with feedback models is generally more \emph{robust} to different query variations. In fact, BM25 with RM3 or Rocchio is competitive with Query2Doc and MuGI Concat — and more effective than Naive Concat — on the BEIR datasets despite not leveraging an LLM. 
\end{itemize}
\section{Conclusion}

In this work, we investigate whether HyDE can be further improved by drawing from the long-standing literature on feedback models for \emph{pseudo}-relevance feedback. Through experiments across 14 datasets, our results demonstrate that incorporating feedback models can improve HyDE's effectiveness by up to 1.4 points (4.2\%) — with 2.2 points (6\%) improvement on low-resource tasks — when compared to the strongest string concatenation baselines.

A potential question is: \emph{why are feedback models better than string concatenation methods?} Our results suggest that improvements come from two sources. The first is that feedback models implicitly filter out noisy terms, keeping only terms that are meaningful within the corpus and assigned high weight within HyDE's generations. The second improvement, we hypothesize, comes from a more stable and linear weighting of query terms and terms in feedback documents via simple parameters like $\alpha$ and $\lambda$ rather than via a string repeat. This may help explain why the average vector, RM3,  and Rocchio feedback models demonstrated stronger robustness on the diverse range of queries present in low-resource BEIR tasks.

Our findings highlight that traditional feedback models should not be overlooked when leveraging methods like HyDE for relevance feedback and in fact they represent a simple way to better utilize the expansions generated by LLMs. 

\section*{Acknowledgments}

This research was supported in part by the Natural Sciences and Engineering Research Council
(NSERC) of Canada. We would like to thank Basit Ali, Yijun Ge, Kevin Wang and Felix Labelle for their thoughtful feedback and contributions. 
\bibliographystyle{ACM-Reference-Format}
\bibliography{main}
\appendix

\end{document}